\documentstyle[twoside,aps,prl,multicol]{revtex}
\begin{document}
\def\B.#1{{\bbox{#1}}}
\def\BC.#1{{\bbox{\cal{#1}}}}

\title{Hamiltonian structure of the Sabra shell model of turbulence:\\
  exact calculation of an anomalous scaling exponent} \author {Victor
  S. L'vov, Evgenii Podivilov$^*$ and Itamar Procaccia}
\address{Department of~~Chemical Physics, The Weizmann Institute of
  Science,
  Rehovot 76100, Israel\\
  $^*$ Institute of Automatization and Electrometry, Academy Science
  of Russia, Novosibirsk 630090, Russia} \maketitle
\begin{abstract}
  We show that the Sabra shell model of turbulence, which was
  introduced recently, displays a Hamiltonian structure for given
  values of the parameters. As a consequence we compute exactly a
  one-parameter family of anomalous scaling exponents associated with
  4th order correlation functions.
\end{abstract}
\pacs{??}
\begin{multicols}{2}
The field of turbulence and turbulent statistics cannot pride itself
on a large number of exact results. One of the best known exact result
is the ``fourth-fifth law" of Kolmogorov, which pertains to the third
order moment of the longitudinal velocity differences, fixing its
scaling exponent $\zeta_3$ to unity\cite{Fri}.  This follows from the
conservation of energy, a quadratic invariant, by the Euler part of
the Navier-Stokes equations. This famous result reappeared also in
simplified toy models of turbulence, like shell models, since the
conservation of a quadratic invariant was built into their
definition\cite{Piss93PFA}.  Nevertheless, the high degree of
simplification involved in the shells model did not lead so far to
additional exact results.

In this Letter we report a discovery of a Hamiltonian structure of the
Sabra shell model of turbulence that was introduced
recently\cite{98LPPPV}.  The Hamiltonian structure exists for given
values of the parameters of the model, and it {\em does not coincide}
with the usual quadratic invariants; in fact it is cubic in the
velocities. There is a Hamiltonian density ${\cal H}_n$ which is {\em
  local} in shell-space, and its existence implies a flux of a local
conserved density. The consequence of the constancy of this flux in
shell-space fixes the value of a fourth order scaling exponent which
is ``anomalous" in the sense that it must involve the existence of a
renormalization scale. This result is both exact and nontrivial,
constituting a kind of ``boundary condition'' on theories of anomalous
scaling in such models.

The Sabra model of turbulence, like all shell models of turbulence,
describes truncated fluid mechanics in wave-number space in which we
keep $N+1$ ``shells".  Denoting the $n$th wave-number component of the
velocity field by $v_n$, $0\le n\le N$, the model may be written as
\begin{eqnarray} \label{sabra}
 \dot v_n &=& - i\big( ak_{n+1}  v_{n+1}v_{n+2}
 -  bk_n v_{n-1}v_{n+1}^*  \\ \nonumber
&& -ck_{n-1} v_{n-1}^*v_{n-2}\big)  -\nu k_n^2  v_n +f_n\,,
\end{eqnarray}
where the coefficients $a$, $b$, and $c$ are real, $\nu$ is the
``viscosity" and $f_n$ is a forcing term, usually limited to acting on
the first shells only. Equation (\ref{sabra}) is equivalent to the
Sabra model suggested in\cite{98LPPPV} after the change of variables:
$v_n=u_n$ for $n$ even and $v_n=-u_n^*$ for $n$ odd.  Like in all
shell models one builds in the conservation of ``energy"
$E=\sum_{n=0}^N|v_n|^2$
in the inviscid limit $\nu\to 0$ by requiring that $a+b+c=0$. The actual
values of the
wave-numbers $k_n$ are determined by $k_0$ and the ``level spacing"
parameter
$\lambda$, $k_n=k_0\lambda^n$. By rescaling we can always choose $a=1$,
leaving
us with only two free parameters in this model, $\lambda$ and
$\epsilon=-b$.
The inviscid and unforced part  of our model then reads
\begin{eqnarray}\label{sabraeps}
\dot v_n&=&-ik_{n+1}\big[v_{n+1}v_{n+2}+\epsilon \lambda^{-1}
 v_{n-1}v_{n+1}^*
\\ \nonumber
&&+(1-\epsilon) \lambda^{-2} v_{n-2}v_{n-1}^*\big] \ .
\end{eqnarray}
Like the popular GOY model \cite{Piss93PFA}, the Sabra model also
conserves
the so-called ``helicity''
\begin{equation}\label{eq:helicity}
  H=\sum_{n=0}^N(\epsilon -1)^{-n}|v_n|^2\ .
\end{equation}
For the special choice of
$\epsilon$ equal to the golden mean $\epsilon_g=(\sqrt{5}-1)/2$,
Eq.(\ref{sabraeps})
reads
\begin{eqnarray}\label{sabragold}
\dot v_n&=&-ik_{n+1}\Big[v_{n+1}v_{n+2}+{\epsilon_g \over\lambda}
v_{n-1}v_{n+1}^*
\\ \nonumber
&&+\Big({\epsilon_g\over \lambda}\Big)^2  v_{n-2}v_{n-1}^*\Big] \ .
\end{eqnarray}
By a direct calculation one can verify that in this case the model
exhibits
a third integral
of motion which we denote as  $W$:
\begin{equation}\label{Ham}
W= \sum_{m=1}^{N-1} \epsilon_g  k_0\left({\lambda\over
\epsilon_g }\right)^n
\left[v^*_{n-1}v_nv_{n+1}+ {\rm c.c.}\right] \ .
\end{equation}

The main
statement of this Letter is that the Sabra model (\ref{sabragold})
with $\epsilon$ chosen at the golden mean realizes
a Hamiltonian structure in the sense that the equations of motion can
be written in a canonical form
\begin{equation}\label{canon}
\dot a_n=-i{\partial {\cal H}\{a_j,a_j^*\}         \over \partial a^*_n}
\,,
\end{equation}
with ${\cal H}\{a_j,a_j^*\} $ being the Hamiltonian function of the set of
pairs of
canonical variables $\{a_j,a_j^*\}$ representing all the degrees of
freedom.
The Hamiltonian $\cal H$ is precisely the
cubic invariant $W$, expressed in terms of canonical variables $a_n$,
$a_n^*$,  which are  related to the velocity as follows:
\begin{equation}
a_n\equiv v_n/\epsilon_g^{n/2} \ . \label{defan}
\end{equation}
According to  (\ref{Ham})   and     (\ref{defan}) the
Hamiltonian has the  form: ${\cal H}=\sum_{m=1}^{N-1} {\cal
  H}_m$  where 
\begin{equation}\label{HHm}
{\cal H}_m=\epsilon_g k_0(\lambda\sqrt{\epsilon_g})^m
\left[a_{m+1}a_ma^*_{m-1}+{\rm  c.c.}
\right] \ .
\end{equation}
The statement is proven by verification: compute $\dot a_n$ according to
the canonical equations (\ref{canon}) 
with the Hamiltonian $ {\cal H}$:
\begin{eqnarray}
  \label{eq:can-a}
  \dot a_n&=&-i {\partial {\cal H}\over \partial a_n^*} =
- i k_0\epsilon (\lambda\sqrt{\epsilon_g})^n
  [\lambda\sqrt{\epsilon_g} a_{n+2}a_{n+1} \\
&&+ a_{n+1}^*a_{n-1} +
  \frac{1}{\lambda\sqrt{\epsilon_g
}} a_{n-1}^*a_{n-2}].   \nonumber
\end{eqnarray}
Using the relationship (\ref{defan}) this equation  coincides with
Eq.~(\ref{sabragold}), {\sl Q.E.D.}

One should note that in the majority of physical systems in the
conservative limit
the Hamiltonian, if it exists, coincides with the energy. Examples abound:
electrodynamics, hydrodynamics,  acoustics, waves on the surface of
fluids,
spin waves
etc, see for example\cite{ZLF}). There are no
regular procedures to find the canonical
variables in terms of the ``natural'' variables of a particular physical
system.
This relationship
may be very nonobvious, like the Clebsch representation of the
velocity field in hydrodynamics.
There are also examples of physical problems with a Hamiltonian structure
in which the Hamiltonian {\em does not} coincide with the energy, like
induced
scattering of electrons
on  ions in strongly non-isothermal plasmas\cite{75ZMR}. The present
example belongs to this class.

The existence of a Hamiltonian
has significant consequences for the solutions
of nonlinear problems. In this Letter we demonstrate one such consequence,
which is  the ability to determine exactly an anomalous exponent.
The locality of ${\cal H}_n$ in the space of shell numbers
means that there exist a current  $J_4(k_n) $  over shells of the mean
value of the``density"
$\langle {\cal H}_n\rangle$. Here the symbol $\langle\dots\rangle$ stands
for an appropriate ensemble average. In order to derive an expression for
$J_4(k_n) $   compute the time derivative of ${\cal H}_n$ (\ref{HHm})
with the help of the canonical equations (\ref{eq:can-a}).  The
right-hand-side of this equation is a combination of 4th-order
correlation functions. The conservation of the Hamiltonian means
 that $ \sum_n\dot{\cal H}_n =0$. This is possible only if
$\langle \dot{\cal H}_n \rangle $
can be written as a {\sl difference} of currents depending on two
neighboring  shells (substituting the divergence in the continuous limit):
\begin{equation}
\langle \dot {\cal H}_n\rangle=J_4(k_{n-1})-J_4(k_n) \ .
\end{equation}
We find
\begin{equation}\label{JF}
J_4(k_n)=-\epsilon_g  k_0^2\Big({\lambda^2\over \epsilon_g}\Big)^n \tilde
F_4(k_n)\ ,
\end{equation}
where $\tilde F_4(k_n)$ is the following combination of 4th order velocity
correlation
functions :
\FL
\begin{eqnarray}\label{deftildF4}
\tilde F_4(k_n) &
\equiv & {\rm Im}[\lambda^2\langle
v_{n-1}^*v_nv_{n+2}v_{n+3}\rangle\\ \nonumber
&& +\epsilon_g \langle
v_{n-2}^*v_{n-1}v_{n+1}v_{n+2}\rangle\nonumber\\&&+\epsilon_g\lambda
\langle v_{n-1}^*v_n^2v_{n+2}^*\rangle+\lambda\langle
v_{n-1}^*v_n^2v_{n+1}^2v_{n+2}\rangle]
\ . \nonumber
\end{eqnarray}
In the steady state $\langle \dot {\cal H}_n\rangle=0$,  implying that
$J_4(k_n)$ must be independent of
$k_n$:             $J_4(k_n)=J_4(k_{n-1})$.
According to (\ref{JF})  we must demand
\begin{equation}
\tilde F_4(k_n)\propto \left({\epsilon_g\over \lambda^2}\right)^n \ ,
\end{equation}
or, looking for a scaling solution $\tilde F_4(k_n)\propto k_n^{-\tilde
\zeta_4}$, we compute
\begin{equation}
\tilde \zeta_4 = \log_\lambda{\Big({
\lambda^2\over\epsilon_g}\Big)}=2+\log_\lambda
{\Big({\sqrt{5}+1\over 2}\Big)}\ .
\end{equation}
This is an exact solution for a one-parameter family of scaling exponents
as a
function
of the level spacing $\lambda$.

We note that
the anomalous exponent $\tilde \zeta_4\ge  2$.   The H\"older inequalities
imply that
the scaling exponent $\zeta_4$ characterizing any of the correlation
functions constituting $\tilde F_4(k_n)$ is smaller or equal to $4/3$.
Denote the scale at which the Hamiltonian invariant is pumped into the
system by $k_{_{\cal H}}$. If this scale is
chosen to coincide with the first shells of the model, the anomalous
exponent found here will always be subleading. This means that the leading
scaling
contributions exactly cancel in the combination of correlation functions
making
up $\tilde F_4(k_n)$ due to the conservation of the Hamiltonian. On the
other hand, if $k_{_{\cal H}}$ is chosen in the bulk of the inertial
interval, and if this has an effect on the region of smaller $k<k_{_{\cal
H}}$,
then in that region the contribution of $\tilde \zeta_4$ will be leading.
This and other consequences of the existence of the Hamiltonian structure
will be investigated further in the near future.

This work was supported in part by the US-Israel Bi-National Science
Foundation, The German-Israeli Foundation, the Basic Research Fund
administered by the Israeli Academy of Science
and the Naftali and Anna Backenroth-Bronicki Fund for
Research in Chaos and Complexity.

\end{multicols}
\end{document}